\documentclass{article}

\begin{document}

\title{Actions for Biconformal Matter}
\author{Andr\'{e} Wehner\footnote{Utah State University Department of Physics, Logan, UT 84322-4415, sllg7@cc.usu.edu}  and James T. Wheeler\footnote{Utah State University Department of Physics, Logan, UT 84322-4415, jwheeler@cc.usu.edu}}
\maketitle

\begin{abstract}
We extend $2n$-dim biconformal gauge theory by including
Lorentz-scalar matter fields of arbitrary conformal weight. For a massless
scalar field of conformal weight zero in a torsion-free biconformal
geometry, the solution is determined by the Einstein equation on an $n$%
-dim submanifold, with the stress-energy tensor of the scalar field
as source. The matter field satisfies the $n$-dim Klein-Gordon
equation.
\end{abstract}

\section{Introduction}

Recently, we developed a new gauge theory of the conformal group, which
solved many of the problems typically associated with scale invariance \cite
{New Conformal Gauging Paper}. In particular, this new class of \textit{%
biconformal geometries} has been shown to resolve the problem of writing
scale-invariant vacuum gravitational actions in arbitrary dimension without
the use of compensating fields \cite{WW}. In the cited work, we wrote the
most general linear vacuum action and completely solved the resulting field
equations subject only to a minimal torsion assumption. We found that all
such solutions were foliated by equivalent $n$-dimensional Ricci-flat
Riemannian spacetimes.

Reference \cite{WW} left an open question: How are matter fields coupled to
biconformal gravity? \textit{A priori}, it is not at all obvious that any
action for biconformal matter permits the same embedded $n$-dimensional
Riemannian structure that occurs for the vacuum case since biconformal
fields are functions of all $2n$-dimensions. Indeed, in the case of standard $n$-dimensional conformal gauging (\cite{Romao+Ferber+Freund}-\cite{Kaku+Townsend2}), we generally require compensating fields to recover the Einstein equation with
matter (see, for example, \cite{Deser}-\cite{van Nieuwenhuizen}). To answer
this question for biconformal space, we extend the results of \cite{WW} by
introducing a set of Klein-Gordon-type fields $\phi ^{m}$ of conformal
weight $m$ into the theory.\footnote{%
Details of the calculations described in this letter can be found at arXiv:
hep-th/0001061.} Using the Killing metric intrinsic to biconformal space, we
write the natural kinetic term in the biconformally covariant derivatives of 
$\phi ^{m}$ and find the resulting gravitationally coupled field equations.
Then, for the case of one scalar field $\phi $ of conformal weight zero, we
solve the new coupled gravitational and scalar field system, under the
assumption of vanishing torsion. At the outset there are two likely
outcomes: Either the presence of the scalar field will destroy the
submanifold structure of the purely gravitational system, or the submanifold
structure will be imposed on the fields. Which of these occurs is the
central issue. We find that, as before, the solutions are foliated by
equivalent $n$-dimensional Riemannian spacetime submanifolds whose
curvatures now satisfy the usual Einstein equations with scalar matter. The
field $\phi ,$ which \textit{a priori }depended on all $2n$ biconformal
coordinates, is completely determined by the $n$ coordinates of the
submanifolds and satisfies the massless Klein-Gordon equation on each
submanifold.

Thus, the new gauging establishes a clear connection between conformal gauge
theory and general relativity with scalar matter, without the use of
compensating fields.

\section{The biconformal inner product and dual}

Full detail on the new gauging of the conformal group $O(n,2)$, $n>2$, is
available in \cite{New Conformal Gauging Paper}. We refer to the connection
components associated with the Lorentz, dilation, translation, and special
conformal transformation generators of $O(n,2)$ as the spin connection, the
Weyl vector, the solder form, and the co-solder form, respectively. We refer
to the corresponding $O(n,2)$ curvature components as the curvature,
dilation, torsion, and co-torsion, respectively.

Biconformal space possesses a natural metric $K^{AB}$ ($A,B=1\ldots n$),
which is obtained when the non-degenerate Killing form of $O(n,2)$ is
restricted to the $2n$-dimensional base space. The Killing metric provides a
natural inner product between two $r$-forms $\mathbf{U}$ and $\mathbf{V}$
defined over biconformal space.

It is also possible to define the biconformal dual of a general $r$-form, $%
^{\ast }\mathbf{V}$, which is a $2n-r$-form of the same weight as $\mathbf{V}
$. Then the term $\mathbf{U\mathbf{\wedge }^{\ast }V=V\wedge }^{\ast }%
\mathbf{U}$ is proportional to both the inner product of $\mathbf{U}$ and $%
\mathbf{V}$ and the scale-invariant volume form of biconformal space, $%
\mathbf{\Phi }$, whose existence we demonstrated in \cite{WW}.

In constructing the biconformal theory of scalar matter we let $\phi ^{m}$
be a set of massless Lorentz-scalar fields of conformal weight $m\in \mathbf{%
Z}$ \cite{Extended conformal paper}, which depend on all $2n$ coordinates.
Then $\mathbf{D}\phi ^{m}$ denotes their biconformally covariant
derivatives. Each such covariant derivative is a one-form which is also of
weight $m$. Since the dual operator preserves the conformal weight, the term 
$^{\ast }\mathbf{D}\phi ^{-m}$ must be of weight $-m$. With appropriate
conventions for signs and combinatorial factors we arrive at the
Weyl-scalar-valued action 
\[
S_{Matter}=\frac{1}{2}\lambda \sum_{m}\mathbf{D}\phi ^{m}\wedge ^{\ast }%
\mathbf{D}\phi ^{-m}=\frac{1}{2}\lambda \frac{(-1)^{n}}{n!^{2}}%
\sum_{m}K^{AB}D_{A}\phi ^{m}D_{B}\phi ^{-m}{\Phi } 
\]
for some constant $\lambda $. Here $D_{A}\phi ^{m}$ denotes the components
of $\mathbf{D}\phi ^{m}$ in the biconformal basis. We make use of the `dual'
form of $S_{Matter}$ when we vary the action with respect to the field and
the Weyl vector, whereas the form of $S_{Matter}$ that explicitly displays
the dependence on the Killing metric proves more useful in varying the
solder and co-solder forms.

\section{The linear scalar action and solution of the field equations}

In a $2n$-dimensional biconformal space the most general Lorentz and
scale-invariant action $S_{Gravity}$ which is linear in the biconformal
curvatures and structural invariants is given in \cite{WW}. For a set of
massless Lorentz scalar fields $\phi ^{m}$ of weight $m$, we now have 
\[
S=S_{Matter}+S_{Gravity}. 
\]

Variation of this action with respect to the scalar fields yields the wave
equation for each $\phi ^{m}$. Variation with respect to the connection
one-forms gives rise to the field equations for the various components of
the biconformal curvatures. Their form is essentially the same as in \cite
{WW}, except that six of the eight gravitational field equations are now
coupled to matter sources. Due to the presence of these sources, it not
obvious that the submanifold character of the general solution for the
curvatures and connections found in \cite{WW} is still valid. We find,
however, that we can reproduce and extend our results from the vacuum case
such that the usual Einstein theory with scalar matter emerges. This is the
central focus of this letter. In the remainder of this section, we review
the main features of the solution to a sufficient degree that we can
highlight those aspects of the solution that are novel. There are four
principal parts to the solution which we discuss in turn.

\smallskip

First, the field equations for the torsion and co-torsion are solved
algebraically, relating the torsion, co-torsion, Weyl vector and now the
matter fields. Here we find that the presence of the matter fields $\phi
^{m} $ influences only the form of the Weyl vector. Even a weak constraint
on the traces of the torsion and co-torsion determines the Weyl vector
completely in terms of the $\phi ^{m}$ and their covariant derivatives,
which couple in such a way that the Weyl vector vanishes unless conjugate
weights, $+m$ and $-m,$ are both present. This result is in contrast with 
\cite{WW}, where it was observed that constraining the torsion to vanish
also forces the Weyl vector to vanish. Thus, while we assumed in \cite{WW}
that the torsion had to be at least ``minimal'' in order not to constrain
the Weyl vector unduly, it now appears that the Weyl vector vanishes unless
there are appropriate matter fields present. Taking this view, we are free
to assume vanishing torsion, although we note that the minimal torsion
assumption would also give the Weyl vector in terms of the matter.

Vanishing torsion implies the above-mentioned weak trace condition on the
co-torsion. Hence, there exists a gauge in which the Weyl vector is given in
terms of covariant derivatives of the fields $\phi ^{m}$. We will explore
such dilational sources further elsewhere. Here it is sufficient to restrict
our attention to the case $m=0.$ Thus, for the remainder of this paper we
restrict ourselves to the case of one scalar field $\phi $ of conformal
weight zero.

\smallskip

Second, with vanishing torsion, we deduce a reduced form for the curvatures
by combining the remaining four field equations with the Bianchi identities
involving the torsion. Each of these curvature equations now carries a
source. It is here that we begin to see that the submanifold structure
predominates even in the presence of matter. Despite the full stress-energy
sources, we find that the curvatures reduce much as in \cite{WW}, with many of
the components necessarily vanishing. As a result, the matter field $\phi ,$
which \textit{a priori }depended on all $2n$ biconformal coordinates, has
vanishing derivatives in half of the $2n$ directions.

\smallskip

Third, we move to the details of the dimensional reduction. Due to the
vanishing torsion, one of the structure equations is in involution in the $n$
solder forms $\mathbf{e}^{a}$. By the Frobenius theorem, we can consistently
set the $\mathbf{e}^{a}$ to zero and obtain a foliation by $n$-dimensional
submanifolds. Because of the reductions in the form of the curvatures
described above, these submanifolds are flat, with each submanifold spanned
by $n$ one-forms, $\mathbf{f}_{a}$. With appropriate gauge choices the
connection for the subsystem can be found in the usual way in terms of the $%
\mathbf{f}_{a}$ and their derivatives $\mathbf{df}_{a}$.

Returning to the full biconformal space, we find that the $\mathbf{df}_{a}$
remain at least linear in $\mathbf{f}_{a},$ so that the $\mathbf{f}_{a}$ are
also involute. We can therefore set $\mathbf{f}_{a}$ to zero to obtain a
second foliation by $n$-dimensional submanifolds with coordinates $x^{\mu }$%
. These submanifolds are spanned by the $\mathbf{e}^{a}$ and have
nonvanishing curvature.

This brings us to our central result: When the field equations for the
curvature are projected to the $\mathbf{f}_{a}=0$ submanifolds, we obtain
the Einstein equations on each submanifold with the usual stress-energy
tensor, given by $x$-derivatives of the matter field. The solution of the $n$%
-dimensional Einstein equation for the solder form and the $n$-dimensional
wave equation for the scalar field (implicit in the Einstein equation)
determine the full $2n$-dimensional biconformal solution. This establishes a
direct connection between general relativity with scalar matter and the more
general structure of biconformal gauge theory with scalar matter.

\smallskip

Finally, we examine the field equation for $\phi (x^{\mu })$, which is the $%
2n$-dimensional wave equation in a null basis, \textit{not }the $n$%
-dimensional wave equation. It is straightforward to check that this
equation is identically satisfied. Thus, $\phi $ is constraint only by the $%
n $-dimensional wave (Klein-Gordon) equation, which emerges in the usual way
as a consequence of the vanishing divergence of the stress-energy tensor.

\section{Conclusions}

We have developed aspects of the theory of scalar matter in biconformal
space. Using the existence of an inner product of $r$-forms and a dual
operator, we constructed an action for a scalar matter field $\phi ^{m}$
coupled to gravity and found the field equations. We solved them for the
case of a scalar field of conformal weight zero in a torsion-free
biconformal geometry. As in the vacuum case, the generic solutions are
foliated by equivalent $n$-dimensional Riemannian spacetime manifolds. The
curvature of each submanifold satisfies the usual Einstein equations with
scalar matter. The scalar field is entirely defined on the submanifold and
satisfies the $n$-dimensional massless Klein-Gordon equation. Together these
two fields determine the entire $2n$-dimensional biconformal space.

\end{document}